\documentclass[sigconf]{acmart}

\AtBeginDocument{%
  }

\copyrightyear{2025}
\acmYear{2025}
\setcopyright{cc}
\setcctype{by}
\acmConference[RecSys '25]{Proceedings of the Nineteenth ACM Conference on Recommender Systems}{September 22--26, 2025}{Prague, Czech Republic}
\acmBooktitle{Proceedings of the Nineteenth ACM Conference on Recommender Systems (RecSys '25), September 22--26, 2025, Prague, Czech Republic}\acmDOI{10.1145/3705328.3748020}
\acmISBN{979-8-4007-1364-4/2025/09}

\usepackage{booktabs}
\usepackage{bm}
\usepackage{xspace}
\usepackage{array}
\usepackage{booktabs}
\usepackage{makecell}
\usepackage{multirow} 

\begin{document}

\title{Just Ask for Music (JAM): Multimodal and Personalized Natural Language Music Recommendation}

\author{Alessandro B. Melchiorre}
\authornote{Both authors contributed equally to this research.}
\email{a.melchiorre@criteo.com}
\affiliation{
  \institution{Johannes Kepler University Linz}
  \city{Linz}
  \country{Austria}
}
\affiliation{
  \institution{Criteo AI Lab}
  \city{Paris}
  \country{France}
}
\author{Elena V. Epure}
\authornotemark[1]
\email{eepure@deezer.com}
\affiliation{%
  \institution{Deezer Research}
  \city{Paris}
  \country{France}
}

\author{Shahed Masoudian}
\email{shahed.masoudian@jku.at}
\affiliation{
  \institution{Johannes Kepler University Linz}
  \city{Linz}
  \country{Austria}
}
\author{Gustavo Escobedo}
\email{gustavo.escobedo@jku.at}
\affiliation{
  \institution{Johannes Kepler University Linz}
  \city{Linz}
  \country{Austria}
}
\author{Anna Hausberger}
\email{anna.hausberger@jku.at}
\affiliation{
  \institution{Johannes Kepler University Linz}
  \city{Linz}
  \country{Austria}
}
\author{Manuel Moussallam}
\email{manuel.moussallam@deezer.com}
\affiliation{%
  \institution{Deezer Research}
  \city{Paris}
  \country{France}
}
\author{Markus Schedl}
\email{markus.schedl@jku.at}
\affiliation{
  \institution{Johannes Kepler University Linz and Linz Institute of Technology}
  \city{Linz}
  \country{Austria}
}

%

\renewcommand{\shortauthors}{Melchiorre et al.}


\newcommand{\am}[1]{\textcolor{magenta}{{\textbf{Alex}: #1}}}
\newcommand{\sh}[1]{\textcolor{orange}{{\textbf{shahed}: #1}}}

\newcommand{\random}{{Random}\xspace}
\newcommand{\pop}{{Pop}\xspace}
\newcommand{\twotower}{{TwoTower}\xspace}
\newcommand{\talkrec}{{TalkRec}\xspace}
\newcommand{\avgjam}{{AvgMixing}\xspace}
\newcommand{\crsjam}{{CrossMixing}\xspace}
\newcommand{\moejam}{{MoEMixing}\xspace}
\newcommand{\jamdata}{{JAMSessions}\xspace}
\newcommand{\jam}{JAM\xspace}
\newcommand{\eg}{e.\,g., }
\newcommand{\ie}{i.\,e., }
\newcommand{\wrt}{w.\,r.\,t., }
\newcommand{\cf}{c.\,f.\, }

\begin{abstract}
Natural language interfaces offer a compelling approach for music recommendation, enabling users to express complex preferences conversationally. While Large Language Models (LLMs) show promise in this direction, their scalability in recommender systems is limited by high costs and latency. Retrieval-based approaches using smaller language models mitigate these issues but often rely on single-modal item representations, overlook long-term user preferences, and require full model retraining, posing challenges for real-world deployment. In this paper, we present \textbf{\jam} (\textbf{J}ust \textbf{A}sk for \textbf{M}usic), a lightweight and intuitive framework for natural language music recommendation. \jam models user–query–item interactions as vector translations in a shared latent space, inspired by knowledge graph embedding methods like TransE. To capture the complexity of music and user intent, \jam aggregates multimodal item features via cross-attention and sparse mixture-of-experts. We also introduce \textbf{\jamdata}, a new dataset of over 100k user–query–item triples with anonymized user/item embeddings, uniquely combining conversational queries and user long-term preferences. Our results show that \jam provides accurate recommendations, produces intuitive representations suitable for practical use cases, and can be easily integrated with existing music recommendation stacks.

\end{abstract}

\ccsdesc[500]{Information systems~Recommender systems}
\ccsdesc[500]{Information systems~Music retrieval}
\ccsdesc[300]{Information systems~Language models}

\keywords{Recommender Systems, Music Recommendation, Multimodality, Conversational Recommendation, Language Models}

\maketitle

\section{Introduction}
\begin{figure}[t!]
    \centering
    \includegraphics[width=0.8\columnwidth]{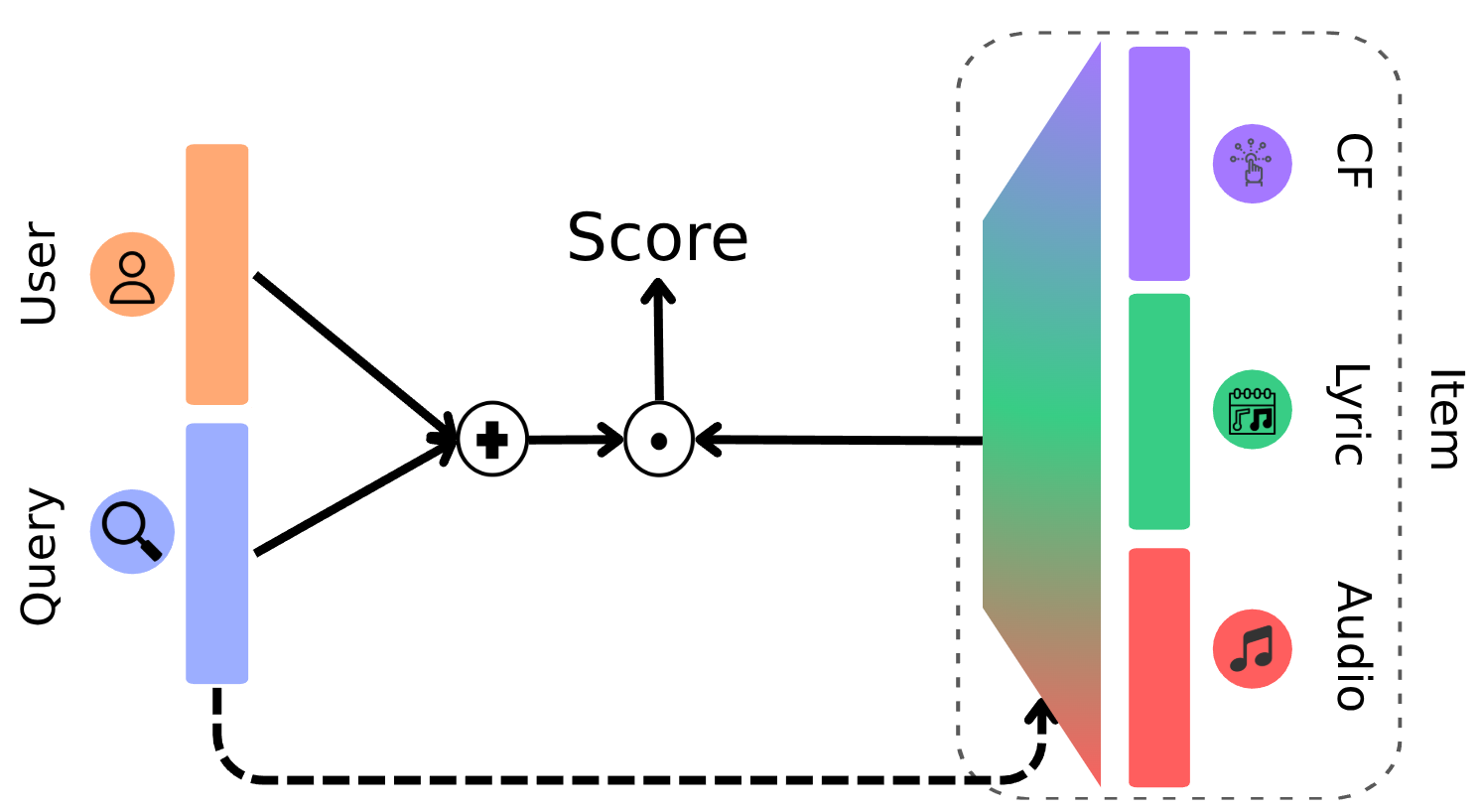}
    \caption{\jam (Just Ask for Music) framework outline.}
    \label{fig:model_outline}
    \Description{This figure shows a sample image used to illustrate the concept.}
\end{figure}

Nowadays music is mostly listened to through streaming platforms like Deezer\footnote{\url{www.deezer.com}} or Spotify\footnote{\url{www.spotify.com}}, which leverage recommendation systems as key components driving engagement. Music consumption is particularly unique due to several factors: users frequently interact with many items in the catalog; feedback is mostly implicit, making it a noisy signal \cite{ferraro2020maximizing}; preferences may shift over time, requiring the disentanglement of short-term explorations from long-term tastes \cite{tran2024transformers}; and listening behavior depends on context and activities \cite{lonsdale2011we, marey2024modeling}. These characteristics add complexity to recommendation models, often requiring multiple building blocks to address each challenge independently.

In this context, music recommender systems with \textit{natural language interfaces} have emerged as a promising way for users to express preferences \cite{jannach2021survey,chen2024large}. This has been driven by advances in natural language processing, particularly with Large Language Models (LLMs) \cite{touvron2023llama,achiam2023gpt}, which have been used as conversational recommenders \cite{lin2025can,zhao2024recommender}. Yet, at an industry scale, relying on LLMs for recommendation has been acknowledged challenging due to high costs required for fine-tuning, suboptimal performance on new music and users, and technical limitations such as limited context size, hallucinations, and significant inference time \cite{oramas2024talking,xu2024prompting}. Consequently, many other works \cite{doh2025talkplay,palumbo2024text2tracks,doh2024music} have framed natural language recommendation as a retrieval task, where queries and items are projected into a shared space learned through contrastive learning \cite{manco2022contrastive}. This allows the use of smaller language models while integrating with established recommender system components like collaborative filtering and content-based methods \cite{oramas2024talking,delcluze2025text2playlist,doh2024music,tekle2024music}.

Despite recent advances, integrating natural language queries with music recommendation systems still faces several limitations. First, many of the proposed solutions rely on \textit{using single sources of information for the item}s (\eg only embedded metadata like track title and artist \cite{chaganty2023beyond}; only tags \cite{epure2024harnessing}; or only audio \cite{manco2022contrastive,doh2024music}). This limits the ability of the existing solutions to effectively align multimodal item representations with complex user queries\footnote{For example, aligning a query like \textit{"love songs"} is inherently challenging if the shared space is built solely on audio features and excludes lyrics.}. Second, while most of the existing solutions account for user preferences expressed in conversation, \textit{long-term preference are seldom considered} \cite{tekle2024music, delcluze2025text2playlist}. This is a critical limitation in music recommendation, where personalization based on historical behavior is central. Third, many pipelines assume \textit{full retraining of the entire model stack}, which is not always practical in real-world systems, where subparts of a recommender system architecture are commonly iterated upon separately \cite{lin2025can, oramas2024talking}. Within this context, we address (\textbf{RQ1}) how to efficiently integrate natural language interfaces into music recommender pipelines with multimodal item representations and short- and long-term user preferences? and (\textbf{RQ2}) which strategies most effectively aggregate such multimodal item data? 

For these purposes, we propose \textbf{\jam} (\textbf{J}ust \textbf{A}sk for \textbf{M}usic, Fig~\ref{fig:model_outline}), a framework that enables natural language interfaces for music recommendation by aligning user preferences expressed in conversation with rich, multimodal item representations, while explicitly accounting for the user's long-term tastes. Inspired by knowledge graph embedding methods such as \textit{TransE}~\cite{bordes2013translating}, we model the personalized recommendation by considering queries ($\textbf{q}$) as translations from the user ($\textbf{u}$) to the item ($\textbf{t}$) representations, thus optimizing a simple equation of the form $\textbf{u} + \textbf{q} \approx \textbf{t}$. 
As multimodality is essential for accurately aligning complex user queries and music items, we explore various strategies to aggregate  heterogeneous sources of item information (collaborative filtering signals, audio, and lyrics) into a single representation, ranging from simple averaging to cross-attention aggregation \cite{vaswani2017attention}, or sparse mixture-of-experts modeling \cite{jacobs1991adaptive}. 
Moreover, we provide a qualitative analysis of the learned latent space of user–query–item interactions, showing how our approach reveals relevant properties.

To train and assess \jam, we introduce \textbf{\jamdata}, \textit{a new real-world dataset of over 100k user-query-item triples}, which includes pre-computed embeddings for users and items. Unlike prior datasets, our release captures both conversational intent and user long-term preference signals, enabling more realistic work on personalized music recommendation via natural language.

In brief, our contributions are: \textbf{(i)} \jam, a lightweight framework that integrates natural language interfaces with multimodal and personalized music recommendation; \textbf{(ii)} an evaluation of multimodal item representation aggregation strategies in this setting; \textbf{(iii)} \jamdata, a new dataset with over 100k user–query–item triples; and \textbf{(iv)} a qualitative analysis of the learned translation-based embedding space.



\section{Dataset}

Building a music recommendation system based on natural language queries and long-term user preferences requires user-query-track data, which is only partially present in existing datasets. The Million Playlist Dataset  \cite{DBLP:conf/recsys/ChenLSZ18}, the Melon Dataset \cite{ferraro2021melon}, and the PlayNTell Dataset \cite{gabbolini-etal-2022-data} provide information on tracks and playlist titles or descriptions, which can be heuristically interpreted as queries, but do not include user aspects. The widely used Million Song Dataset \cite{mcfee2012million} contains user–track interactions, yet lacks any queries. The Conversational Playlist Creation Dataset (CPCD) \cite{chaganty2023beyond} includes music recommendation dialogues between humans; however, its size is limited for training a system.
An alternative strategy involves mining user-generated playlists, such as the 30Music dataset \cite{DBLP:conf/recsys/TurrinQCPC15}, though titles and descriptions are often noisy or repetitive. 

To fill this gap, we present \textbf{\jamdata}, a dataset of \textit{112,337 user-query-item triplets} including \textit{103,752 unique users} and \textit{99,865 unique tracks}. The triplets were sampled from the search logs of a music streaming service over a 1-week period in March 2025. Each data point corresponds to a user entering a query in the search bar and, after exploring the results, landing on an editor-curated playlist they listened to for over 10 minutes. These actions give us insights into the user's short-term intent, so we store the search query (e.g., "sport"), along with the playlist’s title and description (e.g., "Motivation Sports – Get moving with this catchy music selection"), the user, and the playlist tracks relevant to the query.

As user queries are repetitive and rather short we opted for augmenting their context and variability by exploiting the playlist title and descriptions. We use the \textit{DeepSeek-R1-Distill-Qwen-7B}\footnote{\url{https://huggingface.co/deepseek-ai/DeepSeek-R1-Distill-Qwen-7B}} LLM, limiting its output to 20 English words. A two-shot prompt with complex query examples from a separate component of the music streaming service guided the generation.\footnote{Full prompt available in the Appendix.} We conducted an internal quality check of the augmentation with multiple participants, which showed that most augmented queries were accurate, though some erroneous generations still occurred.\footnote{Details are provided in the Appendix.}

The statistics of the \jamdata dataset are shown in the last row of Table~\ref{tab:dataset_stats}. As seen, \jamdata provides large amount of data involving users, queries, and relevant items. The dataset is publicly available on the online repository.
\begin{table}[t]
\begin{tabular}{@{}rcccl@{}}
    Dataset & Queries\footnotemark & Users   & Tracks    \\
    \hline
    MPD~\cite{DBLP:conf/recsys/ChenLSZ18} & 1,000,000 & - & 2,262,292 \\
    Melon~\cite{ferraro2021melon} & 148,826  & - & 649,091 \\
    MSD~\cite{mcfee2012million} & - & 1,019,318 & 384,546 \\
    30Music~\cite{DBLP:conf/recsys/TurrinQCPC15} & 57,561 &  45,167 & 5,675,143  &\\
    CPCD~\cite{DBLP:conf/sigir/ChagantyLZGBR23} & 917 & 917 & 106,736\\
    \hline
    \textbf{\jamdata} & 112,337 & 103,752 & 99,865  \\
    \end{tabular}
    \caption{Statistics of our dataset in comparison to other datasets available.}
    \label{tab:dataset_stats}
\end{table}

\footnotetext{In the absence of an explicit user query, the playlist title and description are assumed to serve as a proxy for the user's intent.}

\section{Methodology}

Let $\mathcal{U}=\{u_i\}^N_{i=1}$ and $\mathcal{T}=\{t_j\}^M_{j=1}$ denote the set of $N$ users and $M$ music items. A user $u$ creates a textual query $q$ (\eg "\textit{sad piano songs}"), which is linked to a collection of music tracks $t_j$ that fulfill it (\eg a playlist of melancholic classical pieces). The goal is to identify the items in $\mathcal{T}$ that best match the query $q$ for the user~$u$, by learning a function that assigns high scores to relevant user-query-item matches. For simplicity, we omit user and item indices.

Each item $t$ in the catalog is represented with multiple modality-specific embeddings $\tilde{\bm{t}}^1, \tilde{\bm{t}}^2, \ldots, \tilde{\bm{t}}^{N_{\text{mod}}}$, each capturing a different aspect of the music track, such as audio, lyrics, or collaborative filtering signals. These representations are typically extracted using pre-trained models \cite{doh2025talkplay,moscati2024multimodal}, each of these models being developed separately. We consider users being represented by a single embedding $\tilde{\bm{u}}$ that reflects their long-term music preferences, \eg their collaborative filtering profile compiled over the item corpus. In contrast, the user's query $q$ captures their short-term preferences or intents expressed in natural language. We use the  \textit{ModernBert-base}\footnote{\url{https://huggingface.co/answerdotai/ModernBERT-base}} text encoder~\cite{warner2024smarter} to obtain a dense representation $\tilde{\bm{q}}$ of the query.

In these settings, we propose \textbf{\jam} (\textbf{J}ust \textbf{A}sk for \textbf{M}usic) —a lightweight framework that seamlessly integrates into existing recommendation ecosystems. We address a common scenario in industry where parts of the recommender system pipeline are already in place, and their outputs are used by downstream components. To operate within this setting, we keep the initial user ($\tilde{\bm{u}}$), query ($\tilde{\bm{q}}$), and item ($\tilde{\bm{t}}^i$) representations \textit{fixed}.

In JAM, the initial representations of users, items, and queries are first projected into a shared latent space of dimensionality $d$ using different encoders, each implemented as 1-layer feed-forward neural network:
$$ \bm{u} = \bm{W}_{\tilde{u}}\tilde{\bm{u}} \quad \bm{q} = \bm{W}_{\tilde{q}}\tilde{\bm{q}} \quad \bm{t}^i = \bm{W}_{\tilde{t}^i}\tilde{\bm{t}}^i \quad \bm{u},\bm{i},\bm{t}^i \in \mathbb{R}^d $$
Inspired by the geometric intuition behind knowledge graph embedding methods such as \textit{TransE}~\cite{bordes2013translating, tran2021hierarchical, he2017translation}, we model personalized recommendation by treating queries $\bm{q}$ as translations from users $\bm{u}$ to items $\bm{t}$, leading to the simple and intuitive formulation:
$$\bm{u} + \bm{q} = \bm{\hat{t}}$$ 
where $\bm{\hat{t}}$ is the aggregated multimodal item representation, discussed below. This formulation enables the learning of a latent space with interesting properties, where the same query translation, \eg \textit{“Something danceable”}, can lead to different item recommendations depending on the user's starting point, and vice versa. We illustrate such cases in Section~\ref{sec:results}.

As each item is associated with multiple modality-specific representations, we explore three different strategies to aggregate these into $\bm{\hat{t}}$ before the matching with the user and query.

\textbf{Averaging (\avgjam).}
A simple strategy to combine the different multimodal representations is averaging them together into a single embedding.
$$\bm{\hat{t}} = \frac{1}{N_{mod}} \sum^{N_{mod}}_i \bm{t}^i $$
While straightforward, it weights all modalities equally, which may be suboptimal in cases where some modalities are more informative or more relevant to the query, \eg \textit{“an upbeat motif”} may depend more heavily on audio features. \\
\textbf{Cross-Attention (\crsjam).}
To dynamically adjust the weighting of the different modalities, we opt for the cross-attention \cite{vaswani2017attention}:
$$\bm{\hat{t}} = \sum^{N_{mod}}_i \alpha(\tilde{\bm{t}}^i , \tilde{\bm{q}})\bm{t}^i $$
which uses the query $\tilde{\bm{q}}$ to compute the scaled dot-product attention over the initial item representations $\tilde{\bm{t}}^i$:
$$ \alpha(\tilde{\bm{t}}^i , \tilde{\bm{q}}) = Softmax\left( \frac{(\bm{W}^{key}_{\tilde{t}^i} \tilde{\bm{t}}^i)^\top (\bm{W}^{query}_{\tilde{q}} \tilde{\bm{q}})}{\sqrt{d}} \right)$$
This strategy allows the query to dynamically weight the contribution of each modality, focusing on the most relevant aspects of the item representations based on the user's short-term preference.\\
\textbf{Sparse Mixture of Experts (\moejam).}
Mixture of Experts \cite{jacobs1991adaptive} combines the representations from different modalities, considering each item modality as an expert. In particular, we leverage the Noisy Top-K gating proposed by \citet{shazeer2017outrageously}
$$\bm{\hat{t}} = \sum^{N_{mod}}_i \alpha(\tilde{\bm{t}}^i , \tilde{\bm{q}})\bm{t}^i $$
$$\alpha(\tilde{\bm{t}}^i , \tilde{\bm{q}}) = Sotfmax(KeepTopK(H(\tilde{\bm{t}}^i , \tilde{\bm{q}}))$$
$$H(\tilde{\bm{t}}^i , \tilde{\bm{q}}) = \bm{x}^{gate}+ StandardNormal()\cdot Softplus(\bm{x}^{noise})$$
$$\bm{x}^{gate} = (\bm{W}^{gate}_{\tilde{t}^i} \tilde{\bm{t}}^i)^\top (\bm{W}^{gate}_{\tilde{q}} \tilde{\bm{q}}) \qquad \bm{x}^{noise} = (\bm{W}^{noise}_{\tilde{t}^i} \tilde{\bm{t}}^i)^\top (\bm{W}^{noise}_{\tilde{q}} \tilde{\bm{q}})$$
where $KeepTopK$ places $-\infty$ values for modalities not in the Top-K. Effectively, this formulation constraints the model to only leverage up to K item modalities to answer a single query. The value of K is a hyperparameter, set to 2 unless otherwise specified.

Finally, we train \jam and its model variants using a dataset of $(u,q,t)$ triplets. For each positive triplet satisfying $\bm{u} +\bm{q} = \bm{t}$, we sample negative items $t^{neg}$ that are not relevant to the user's query. Following the \textit{TransE}~\cite{bordes2013translating} approach, we maximize the similarity for positive triplets while minimizing it for the negative ones:
$$\mathcal{L} = - \sum_{(u,q,t)\in \mathcal{D}} \sum_{t^{neg}} \log \sigma(sim(\bm{u}+\bm{q},\bm{\hat{t}}) - sim(\bm{u}+\bm{q},\bm{\hat{t}}^{neg})) $$
We compute similarity between the predicted and target items using the standard dot product operation. This formulation effectively corresponds to the standard BPR recommendation loss~\cite{rendle2012bpr}.

\section{Experimental Setup}
\label{sec:setup}

Our experiments are based on real-world data from Deezer, an international music streaming service.
Users are represented via collaborative filtering embeddings, while items are tracks, described through three modalities: audio, lyrics, and collaborative filtering. Audio representations are extracted from the audio signal via contrastive learning similarly to \citet{meseguer2024experimental}; lyrics embeddings are created using the \textit{multilingual-e5-base} \cite{wang2024multilingual} model on the full lyrics; while user/item CF embeddings are derived from factorizing a user-track interaction matrix, weighted by listening recency and frequency. We split the data chronologically by timestamp: queries from the last day form the test set, those from the previous day serve as the validation set, and the remaining data is used for training.

We compare the accuracy of our approach against two relevant baselines in multimodal music retrieval. The first is \citet{oramas2024talking} (\textbf{\talkrec}), which embeds the various multimodal representations into a shared latent space and applies contrastive learning across pairwise modality combinations. In this setup, the query is treated as an additional modality, while the user is not explicitly modeled. As a second baseline, we adopt the widely used Two-Tower model from recommendation literature \cite{tekle2024music,de2024personalized,covington2016deep} 
(\textbf{\twotower}), where the different item representations are concatenated and passed through several neural layers. The user-side is modeled similarly, but no query information is incorporated. Note that the baselines correspond to modeling approaches where either the user or the query is not explicitly considered. Lastly, we also include simple baselines such as random item (\textbf{\random}) and most popular item (\textbf{\pop}) recommendation.

To evaluate the models, we use: \textit{Recall} (the proportion of relevant items successfully retrieved) and Normalized Discounted Cumulative Gain (\textit{NDCG}), which emphasizes high rankings of relevant items in the result list. We report these metrics at cut-off thresholds of 10 and 100 to reflect different user browsing depths.

We train each model for 50 epochs with the AdamW optimizer~\cite{loshchilov2017decoupled} and a cosine annealing learning rate scheduler~\cite{loshchilov2016sgdr}. We fix the batch size to 512 and sample 4 negative items per positive instance. Early-stopping is applied if the NDCG@10 on the validation set does not improve for 10 consecutive epochs. We tune the embedding dimension $d$ and the learning rate for all baselines.\footnote{Details on the hyperparameter search and selected values are provided in the appendix} The best model on the validation set is evaluated on the test set. All experiments are repeated with three random seeds, and we report the mean and standard deviation of the metrics. Code and resources are available at \url{https://github.com/hcai-mms/jam}.

\section{Results}
\label{sec:results}
\begin{table}[t]
\centering
\begin{tabular}{@{}>{\centering\arraybackslash}m{1mm}
                >{\centering\arraybackslash}m{25mm}|
                >{\centering\arraybackslash}m{.5mm}
                >{\centering\arraybackslash}m{1.2mm}|
                >{\centering\arraybackslash}m{8mm}
                >{\centering\arraybackslash}m{9mm}|
                >{\centering\arraybackslash}m{8mm}
                >{\centering\arraybackslash}m{8mm}@{}}
                        &                       &     &     & \multicolumn{2}{c|}{Recall} & \multicolumn{2}{c}{NDCG}                           \\ 
                        &                       & $q$   & $u$   & \centering @10 & @100 & @10 & @100 \\ \midrule
                        & \random               & x   & x   & $.000_{.000}$ & $.001_{.000}$ & $.001_{.000}$ & $.001_{.000}$ \\
                        & \pop                  & x   & x   & $.012_{.000}$ & $.084_{.000}$ & $.073_{.000}$ & $.079_{.000}$ \\
                        & \talkrec              & \checkmark & x   & $.041_{.000}$ & $.159_{.000}$ & $.152_{.000}$ & $.144_{.000}$ \\
                        & \twotower             & x   & \checkmark & $.024_{.002}$ & $.136_{.007}$ & $.110_{.003}$ & $.118_{.006}$ \\ \midrule
\multirow{4}{*}{\shortstack[c]{\rotatebox[origin=c]{90}{\textbf{JAM}}}}
                        & \avgjam               & \checkmark & \checkmark & $.072_{.003}$ & $.313_{.013}$ & $.258_{.008}$ & $.274_{.011}$ \\
                        & \crsjam               & \checkmark & \checkmark & $\bm{.086}_{\bm{.002}}$ & $\bm{.371}_{\bm{.005}}$ & $\bm{.311}_{\bm{.001}}$ & $\bm{.327}_{\bm{.003}}$ \\
                        & \moejam ($K=2$)       & \checkmark & \checkmark & $.048_{.002}$ & $.252_{.006}$ & $.180_{.003}$ & $.211_{.001}$ \\
                        & \moejam ($K=1$)       & \checkmark & \checkmark & $.036_{.009}$ & $.165_{.053}$ & $.128_{.039}$ & $.142_{.045}$ \\
\bottomrule
\end{tabular}
\caption{Recall and NDCG (@10, @100) averages on the test set. Subscripts indicate standard deviation. Columns $q$ and $u$ indicate whether query and user representations are used.}
\label{tab:results}
\end{table}

Table~\ref{tab:results} reports the results on the test set as the average of three random seeds, with subscripts indicating the standard deviation.

Addressing \textbf{RQ1}, all \jam variants achieve higher accuracy than the baselines across both metrics and thresholds. Each baseline captures only a partial view of user intent: \twotower includes long-term preferences via the user embedding but ignores the short-term query, while \talkrec incorporates the query but lacks long-term user information. This results in a reasonable drop in accuracy for both. Among the two, \talkrec— which explicitly models the natural language query—performs better.

Addressing \textbf{RQ2}, among the different multimodal aggregation strategies in \jam, \crsjam consistently achieves the best performance, followed by \avgjam and \moejam. Averaging modalities, as done in \avgjam, proves to be a simple yet effective approach for integrating item representations. However, \crsjam further improves performance by using cross-attention to dynamically reweight representations based on their semantic relevance to the query. In contrast, \moejam— which sparsifies activations across experts/modalities—shows a drop in accuracy compared to the other methods. This drop is more pronounced for $K=1$, suggesting that all modalities contribute valuable information for addressing the query. 
\begin{figure}[thb]
    \centering
    \centering
    \includegraphics[width=0.9\columnwidth]{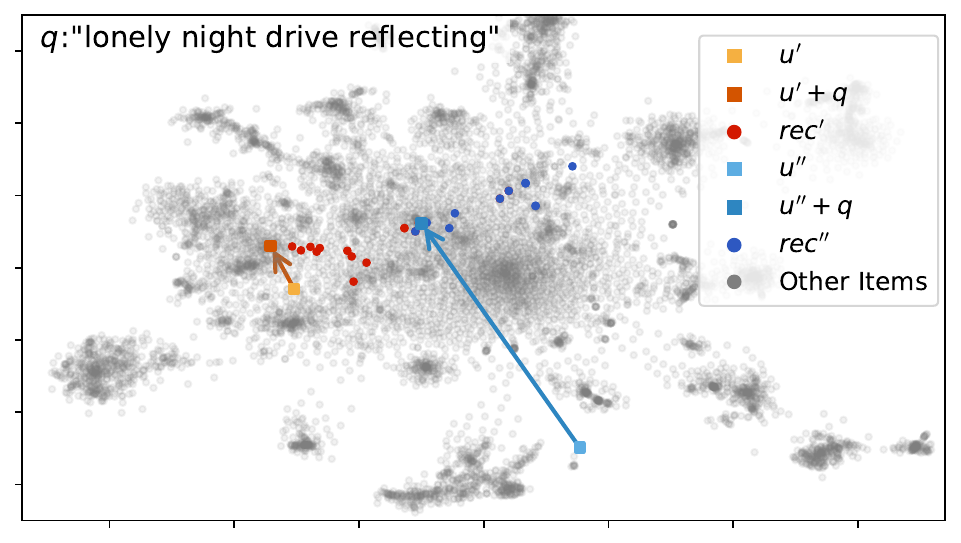}
    \caption{TSNE of item embeddings, two users (u', u"), their translations under the same query (q), and recommendations (rec', rec"). Equivalent translations in $\mathbb{R}^d$ may appear different in 2D.}
    \Description{TSNE of item embeddings, two users (u', u"), their translations under the same query (q), and recommendations (rec', rec"). Equivalent translations in $\mathbb{R}^d$ may appear different in 2D.}

    \label{fig:translation}
\end{figure}

Analysis of MoEMixing’s top-K activations and the highest attention weights in \crsjam indicates that collaborative filtering signals contribute most, likely due to the use of CF embeddings as the main representations of the user. Since user and item CF embeddings are pre-computed jointly, this initialization may bias the model towards favoring this modality. Future work could explore enriching user representations with summaries of user taste to better assess the impact of each modality.
\renewcommand{\arraystretch}{1.35}

\begin{table}[t]
\centering
{\footnotesize
\begin{tabular}{p{0.47\columnwidth}| p{0.47\columnwidth}}
\multicolumn{1}{c|}{$u' + q: \text{"partying like crazy"}$} & \multicolumn{1}{c}{$u'' + q: \text{"partying like crazy"}$} \\ 

\midrule

\parbox[c]{2.8cm}{\textit{Explodiert} by \textit{Harris \& Ford}} \hfill \textbf{Dance} & \parbox[c]{2.6cm}{\vspace{0.1cm}\textit{Ta Hi / Marcha do Remandor}
by \textit{Banda Rio Ipanema} \vspace{0.1cm}} \hfill \textbf{Marchinha}\\


\vspace{0.2em}\\[-0.5em]
\parbox[c]{2.8cm}{\vspace{0.1cm}\textit{Paradies} by \textit{Anstandslos \& Durchgeknallt} \vspace{0.1cm}} \hfill  \textbf{Dance} & \parbox[c]{3.3cm}{\textit{Diga Que Valeu} by \textit{Bell Marques}} \hfill \textbf{Axé}\\

\vspace{0.2em}\\[-0.5em]
\parbox[c]{2.8cm}{\vspace{0.1cm}\textit{HERZ MACHT BAMM} by \textit{Tream}\vspace{0.1cm}} \hfill  \textbf{Dance} & \parbox[c]{3.3cm}{\textit{Ara Ketu Bom Demais} by \textit{Ara Ketu}}  \hfill \textbf{Axé}\\

\hline
\\
\multicolumn{1}{c|}{$u + q': \text{"cartoon music for kids"}$} & \multicolumn{1}{c}{$u + q'': \text{"feeling lonely and sad"}$} \\ 

\hline

\parbox[c]{2.5cm}{\vspace{0.1cm}\textit{Les Aristochats} by \textit{Maurice Chevalier}\vspace{0.1cm}} \hfill  \textbf{Soundtrack} & \parbox[c]{2.5cm}{\textit{What was I made for?} by \textit{Billie Eilish}}  \hfill\textbf{Alternative}\\

\vspace{0.2em}\\[-0.5em]

\parbox[c]{2.6cm}{\vspace{0.1cm}\textit{Tout le monde veut devenir un cat} by \textit{José Germain}\vspace{0.1cm}} \hfill  \textbf{Children} & \parbox[c]{2.5cm}{\textit{State Lines} by \textit{Novo Amor}} \hfill \textbf{Alternative}\\

\vspace{0.2em}\\[-0.5em]

\parbox[c]{2.6cm}{\vspace{0.1cm}\textit{Baby Shark} by \textit{Pinkfong en Français} \vspace{0.1cm}} \hfill  \textbf{Children} & \parbox[c]{2.5cm}{\textit{Oh Love} by \textit{Tomo}}  \hfill \textbf{Folk}\\

\end{tabular}
}
\caption{Top 3 recommendations with (top) same query for different users and (bottom) same user with different queries.}
\label{tab:combined}
\end{table}

Considering the most performant model (\crsjam), we qualitatively assess the effect of modeling queries as translations in the user-item space. Fig.~\ref{fig:translation} shows the projected latent space of multimodal item embeddings, showing two users and their respective translations under the same query \textit{"lonely night drive reflecting"}. With the same query, the two users are translated toward different regions of the space, resulting in personalized recommendations. 

Another example is shown in Tab.~\ref{tab:combined}, which presents top-3 recommendations in two scenarios. In the top half, two users ($u'$ and $u''$) issue the same query (\textit{"partying like crazy"}), but receive different results: $u'$ gets German and Austrian dance tracks, while $u''$ is recommended Brazilian axé and marchinha music. In the bottom half, a single user $u$ issues two queries. The first (\textit{"cartoon music for kids"}) returns tracks from children's movies like Disney’s \textit{The Aristocats} and \textit{Coco}, while the second (\textit{"feeling lonely and sad"}) yields melancholic alternative or folk music, often by solo artists.

The results demonstrate the potential of our approach to deliver personalized music recommendations from natural language queries. A limitation of our approach arises with artist-specific requests (\eg \textit{"Coldplay best songs"}), where unrelated artists may be recommended. This likely stems from data sparsity and the semantic gap between artist names and the available modalities~\cite{gabbolini-etal-2022-data}. Incorporating artist-level signals, such as dedicated embeddings, could help. Still, our work showcases the strength of query-based translation models in navigating multimodal item spaces and tailoring recommendations to diverse user intents.

\section{Conclusion and Future Work}
\balance
We present \jam, a lightweight framework for natural language music recommendation that models queries as translations in a user–item space. By aggregating multimodal item features, long-term user preferences, and textual queries, \jam enables expressive and personalized recommendations while remaining compatible with existing recommendation stacks. We also release \jamdata, a dataset of over 100k user–query–item triples with pre-computed user/item embeddings. Future work includes enriching user profiles with multimodal features and addressing artist-specific queries by incorporating artist embeddings to mitigate sparsity issues.

\begin{acks}
This research was funded, in whole or in part, by the Austrian Science Fund (FWF) under the following grants: \url{https://doi.org/10.55776/COE12}, \url{https://doi.org/10.55776/DFH23}, and \url{https://doi.org/10.55776/P36413}.
The authors thank Aurelien Herault and Viet Anh Tran\footnote{https://research.deezer.com} for their valuable feedback on this work.    
\end{acks} 


\bibliographystyle{ACM-Reference-Format}
\bibliography{sample-base}


\begin{thebibliography}{41}


\ifx \showCODEN    \undefined \def \showCODEN     #1{\unskip}     \fi
\ifx \showISBNx    \undefined \def \showISBNx     #1{\unskip}     \fi
\ifx \showISBNxiii \undefined \def \showISBNxiii  #1{\unskip}     \fi
\ifx \showISSN     \undefined \def \showISSN      #1{\unskip}     \fi
\ifx \showLCCN     \undefined \def \showLCCN      #1{\unskip}     \fi
\ifx \shownote     \undefined \def \shownote      #1{#1}          \fi
\ifx \showarticletitle \undefined \def \showarticletitle #1{#1}   \fi
\ifx \showURL      \undefined \def \showURL       {\relax}        \fi
\providecommand\bibfield[2]{#2}
\providecommand\bibinfo[2]{#2}
\providecommand\natexlab[1]{#1}
\providecommand\showeprint[2][]{arXiv:#2}

\bibitem[Achiam et~al\mbox{.}(2023)]%
        {achiam2023gpt}
\bibfield{author}{\bibinfo{person}{Josh Achiam}, \bibinfo{person}{Steven Adler}, \bibinfo{person}{Sandhini Agarwal}, \bibinfo{person}{Lama Ahmad}, \bibinfo{person}{Ilge Akkaya}, \bibinfo{person}{Florencia~Leoni Aleman}, \bibinfo{person}{Diogo Almeida}, \bibinfo{person}{Janko Altenschmidt}, \bibinfo{person}{Sam Altman}, \bibinfo{person}{Shyamal Anadkat}, {et~al\mbox{.}}} \bibinfo{year}{2023}\natexlab{}.
\newblock \bibinfo{title}{Gpt-4 technical report}.
\newblock


\bibitem[Bordes et~al\mbox{.}(2013)]%
        {bordes2013translating}
\bibfield{author}{\bibinfo{person}{Antoine Bordes}, \bibinfo{person}{Nicolas Usunier}, \bibinfo{person}{Alberto Garcia-Dur\'{a}n}, \bibinfo{person}{Jason Weston}, {and} \bibinfo{person}{Oksana Yakhnenko}.} \bibinfo{year}{2013}\natexlab{}.
\newblock \showarticletitle{Translating embeddings for modeling multi-relational data}. In \bibinfo{booktitle}{\emph{Proceedings of the 27th International Conference on Neural Information Processing Systems - Volume 2}} (Lake Tahoe, Nevada) \emph{(\bibinfo{series}{NIPS'13})}. \bibinfo{publisher}{Curran Associates Inc.}, \bibinfo{address}{Red Hook, NY, USA}, \bibinfo{pages}{2787–2795}.
\newblock


\bibitem[Chaganty et~al\mbox{.}(2023a)]%
        {chaganty2023beyond}
\bibfield{author}{\bibinfo{person}{Arun~Tejasvi Chaganty}, \bibinfo{person}{Megan Leszczynski}, \bibinfo{person}{Shu Zhang}, \bibinfo{person}{Ravi Ganti}, \bibinfo{person}{Krisztian Balog}, {and} \bibinfo{person}{Filip Radlinski}.} \bibinfo{year}{2023}\natexlab{a}.
\newblock \showarticletitle{Beyond single items: Exploring user preferences in item sets with the conversational playlist curation dataset}. In \bibinfo{booktitle}{\emph{Proceedings of the 46th International ACM SIGIR Conference on Research and Development in Information Retrieval}}. \bibinfo{pages}{2754--2764}.
\newblock


\bibitem[Chaganty et~al\mbox{.}(2023b)]%
        {DBLP:conf/sigir/ChagantyLZGBR23}
\bibfield{author}{\bibinfo{person}{Arun~Tejasvi Chaganty}, \bibinfo{person}{Megan Leszczynski}, \bibinfo{person}{Shu Zhang}, \bibinfo{person}{Ravi Ganti}, \bibinfo{person}{Krisztian Balog}, {and} \bibinfo{person}{Filip Radlinski}.} \bibinfo{year}{2023}\natexlab{b}.
\newblock \showarticletitle{Beyond Single Items: Exploring User Preferences in Item Sets with the Conversational Playlist Curation Dataset}. In \bibinfo{booktitle}{\emph{Proceedings of the 46th International {ACM} {SIGIR} Conference on Research and Development in Information Retrieval, {SIGIR} 2023, Taipei, Taiwan, July 23-27, 2023}}. \bibinfo{publisher}{{ACM}}, \bibinfo{pages}{2754--2764}.
\newblock
\href{https://doi.org/10.1145/3539618.3591881}{doi:\nolinkurl{10.1145/3539618.3591881}}


\bibitem[Chen et~al\mbox{.}(2018)]%
        {DBLP:conf/recsys/ChenLSZ18}
\bibfield{author}{\bibinfo{person}{Ching{-}Wei Chen}, \bibinfo{person}{Paul Lamere}, \bibinfo{person}{Markus Schedl}, {and} \bibinfo{person}{Hamed Zamani}.} \bibinfo{year}{2018}\natexlab{}.
\newblock \showarticletitle{Recsys challenge 2018: automatic music playlist continuation}. In \bibinfo{booktitle}{\emph{Proceedings of the 12th {ACM} Conference on Recommender Systems, RecSys 2018, Vancouver, BC, Canada, October 2-7, 2018}}. \bibinfo{publisher}{{ACM}}, \bibinfo{pages}{527--528}.
\newblock
\href{https://doi.org/10.1145/3240323.3240342}{doi:\nolinkurl{10.1145/3240323.3240342}}


\bibitem[Chen et~al\mbox{.}(2024)]%
        {chen2024large}
\bibfield{author}{\bibinfo{person}{Jin Chen}, \bibinfo{person}{Zheng Liu}, \bibinfo{person}{Xu Huang}, \bibinfo{person}{Chenwang Wu}, \bibinfo{person}{Qi Liu}, \bibinfo{person}{Gangwei Jiang}, \bibinfo{person}{Yuanhao Pu}, \bibinfo{person}{Yuxuan Lei}, \bibinfo{person}{Xiaolong Chen}, \bibinfo{person}{Xingmei Wang}, {et~al\mbox{.}}} \bibinfo{year}{2024}\natexlab{}.
\newblock \showarticletitle{When large language models meet personalization: Perspectives of challenges and opportunities}.
\newblock \bibinfo{journal}{\emph{World Wide Web}} \bibinfo{volume}{27}, \bibinfo{number}{4} (\bibinfo{year}{2024}), \bibinfo{pages}{42}.
\newblock


\bibitem[Covington et~al\mbox{.}(2016)]%
        {covington2016deep}
\bibfield{author}{\bibinfo{person}{Paul Covington}, \bibinfo{person}{Jay Adams}, {and} \bibinfo{person}{Emre Sargin}.} \bibinfo{year}{2016}\natexlab{}.
\newblock \showarticletitle{Deep Neural Networks for YouTube Recommendations}. In \bibinfo{booktitle}{\emph{Proceedings of the 10th ACM Conference on Recommender Systems}} (Boston, Massachusetts, USA) \emph{(\bibinfo{series}{RecSys '16})}. \bibinfo{publisher}{Association for Computing Machinery}, \bibinfo{address}{New York, NY, USA}, \bibinfo{pages}{191–198}.
\newblock
\showISBNx{9781450340359}
\href{https://doi.org/10.1145/2959100.2959190}{doi:\nolinkurl{10.1145/2959100.2959190}}


\bibitem[De~Nadai et~al\mbox{.}(2024)]%
        {de2024personalized}
\bibfield{author}{\bibinfo{person}{Marco De~Nadai}, \bibinfo{person}{Francesco Fabbri}, \bibinfo{person}{Paul Gigioli}, \bibinfo{person}{Alice Wang}, \bibinfo{person}{Ang Li}, \bibinfo{person}{Fabrizio Silvestri}, \bibinfo{person}{Laura Kim}, \bibinfo{person}{Shawn Lin}, \bibinfo{person}{Vladan Radosavljevic}, \bibinfo{person}{Sandeep Ghael}, {et~al\mbox{.}}} \bibinfo{year}{2024}\natexlab{}.
\newblock \showarticletitle{Personalized audiobook recommendations at spotify through graph neural networks}. In \bibinfo{booktitle}{\emph{Companion Proceedings of the ACM Web Conference 2024}}. \bibinfo{pages}{403--412}.
\newblock


\bibitem[Delcluze et~al\mbox{.}(2025)]%
        {delcluze2025text2playlist}
\bibfield{author}{\bibinfo{person}{Mathieu Delcluze}, \bibinfo{person}{Antoine Khoury}, \bibinfo{person}{Cl{\'e}mence Vast}, \bibinfo{person}{Valerio Arnaudo}, \bibinfo{person}{L{\'e}a Briand}, \bibinfo{person}{Walid Bendada}, {and} \bibinfo{person}{Thomas Bouab{\c{c}}a}.} \bibinfo{year}{2025}\natexlab{}.
\newblock \showarticletitle{Text2Playlist: Generating Personalized Playlists from Text on Deezer}. In \bibinfo{booktitle}{\emph{The 47th European Conference on Information Retrieval (ECIR 2025)}}.
\newblock


\bibitem[Doh et~al\mbox{.}(2024)]%
        {doh2024music}
\bibfield{author}{\bibinfo{person}{SeungHeon Doh}, \bibinfo{person}{Keunwoo Choi}, \bibinfo{person}{Daeyong Kwon}, \bibinfo{person}{Taesu Kim}, {and} \bibinfo{person}{Juhan Nam}.} \bibinfo{year}{2024}\natexlab{}.
\newblock \showarticletitle{Music Discovery Dialogue Generation Using Human Intent Analysis and Large Language Models}. In \bibinfo{booktitle}{\emph{Proceedings of the 25th International Society for Music Information Retrieval Conference}}. \bibinfo{pages}{946--953}.
\newblock


\bibitem[Doh et~al\mbox{.}(2025)]%
        {doh2025talkplay}
\bibfield{author}{\bibinfo{person}{Seungheon Doh}, \bibinfo{person}{Keunwoo Choi}, {and} \bibinfo{person}{Juhan Nam}.} \bibinfo{year}{2025}\natexlab{}.
\newblock \showarticletitle{TALKPLAY: Multimodal Music Recommendation with Large Language Models}.
\newblock \bibinfo{journal}{\emph{arXiv preprint arXiv:2502.13713}} (\bibinfo{year}{2025}).
\newblock


\bibitem[Epure et~al\mbox{.}(2024)]%
        {epure2024harnessing}
\bibfield{author}{\bibinfo{person}{Elena~V Epure}, \bibinfo{person}{Gabriel Meseguer-Brocal}, \bibinfo{person}{Darius Afchar}, {and} \bibinfo{person}{Romain Hennequin}.} \bibinfo{year}{2024}\natexlab{}.
\newblock \showarticletitle{Harnessing High-Level Song Descriptors towards Natural Language-Based Music Recommendation}. In \bibinfo{booktitle}{\emph{Proceedings of the 3rd Workshop on NLP for Music and Audio (NLP4MusA)}}. \bibinfo{pages}{17--24}.
\newblock


\bibitem[Ferraro et~al\mbox{.}(2021)]%
        {ferraro2021melon}
\bibfield{author}{\bibinfo{person}{Andres Ferraro}, \bibinfo{person}{Yuntae Kim}, \bibinfo{person}{Soohyeon Lee}, \bibinfo{person}{Biho Kim}, \bibinfo{person}{Namjun Jo}, \bibinfo{person}{Semi Lim}, \bibinfo{person}{Suyon Lim}, \bibinfo{person}{Jungtaek Jang}, \bibinfo{person}{Sehwan Kim}, \bibinfo{person}{Xavier Serra}, {and} \bibinfo{person}{Dmitry Bogdanov}.} \bibinfo{year}{2021}\natexlab{}.
\newblock \showarticletitle{Melon Playlist Dataset: a public dataset for audio-based playlist generation and music tagging}. In \bibinfo{booktitle}{\emph{International Conference on Acoustics, Speech and Signal Processing (ICASSP 2021)}}.
\newblock


\bibitem[Ferraro et~al\mbox{.}(2020)]%
        {ferraro2020maximizing}
\bibfield{author}{\bibinfo{person}{Andres Ferraro}, \bibinfo{person}{Sergio Oramas}, \bibinfo{person}{Massimo Quadrana}, {and} \bibinfo{person}{Xavier Serra}.} \bibinfo{year}{2020}\natexlab{}.
\newblock \showarticletitle{Maximizing the engagement: exploring new signals of implicit feedback in music recommendations}. In \bibinfo{booktitle}{\emph{Proceedings of the Workshops on Recommendation in Complex Scenarios and the Impact of Recommender Systems co-located with 14th ACM Conference on Recommender Systems (RecSys 2020)}}. CEUR Workshop Proceedings.
\newblock


\bibitem[Gabbolini et~al\mbox{.}(2022)]%
        {gabbolini-etal-2022-data}
\bibfield{author}{\bibinfo{person}{Giovanni Gabbolini}, \bibinfo{person}{Romain Hennequin}, {and} \bibinfo{person}{Elena Epure}.} \bibinfo{year}{2022}\natexlab{}.
\newblock \showarticletitle{Data-Efficient Playlist Captioning With Musical and Linguistic Knowledge}. In \bibinfo{booktitle}{\emph{Proceedings of the 2022 Conference on Empirical Methods in Natural Language Processing}}. \bibinfo{publisher}{Association for Computational Linguistics}, \bibinfo{address}{Abu Dhabi, United Arab Emirates}, \bibinfo{pages}{11401--11415}.
\newblock
\href{https://doi.org/10.18653/v1/2022.emnlp-main.784}{doi:\nolinkurl{10.18653/v1/2022.emnlp-main.784}}


\bibitem[Ganh\"{o}r et~al\mbox{.}(2024)]%
        {moscati2024multimodal}
\bibfield{author}{\bibinfo{person}{Christian Ganh\"{o}r}, \bibinfo{person}{Marta Moscati}, \bibinfo{person}{Anna Hausberger}, \bibinfo{person}{Shah Nawaz}, {and} \bibinfo{person}{Markus Schedl}.} \bibinfo{year}{2024}\natexlab{}.
\newblock \showarticletitle{A Multimodal Single-Branch Embedding Network for Recommendation in Cold-Start and Missing Modality Scenarios}. In \bibinfo{booktitle}{\emph{Proceedings of the 18th ACM Conference on Recommender Systems}} (Bari, Italy) \emph{(\bibinfo{series}{RecSys '24})}. \bibinfo{publisher}{Association for Computing Machinery}, \bibinfo{address}{New York, NY, USA}, \bibinfo{pages}{380–390}.
\newblock
\showISBNx{9798400705052}
\href{https://doi.org/10.1145/3640457.3688138}{doi:\nolinkurl{10.1145/3640457.3688138}}


\bibitem[He et~al\mbox{.}(2017)]%
        {he2017translation}
\bibfield{author}{\bibinfo{person}{Ruining He}, \bibinfo{person}{Wang-Cheng Kang}, {and} \bibinfo{person}{Julian McAuley}.} \bibinfo{year}{2017}\natexlab{}.
\newblock \showarticletitle{Translation-based recommendation}. In \bibinfo{booktitle}{\emph{Proceedings of the eleventh ACM conference on recommender systems}}. \bibinfo{pages}{161--169}.
\newblock


\bibitem[Jacobs et~al\mbox{.}(1991)]%
        {jacobs1991adaptive}
\bibfield{author}{\bibinfo{person}{Robert~A Jacobs}, \bibinfo{person}{Michael~I Jordan}, \bibinfo{person}{Steven~J Nowlan}, {and} \bibinfo{person}{Geoffrey~E Hinton}.} \bibinfo{year}{1991}\natexlab{}.
\newblock \showarticletitle{Adaptive mixtures of local experts}.
\newblock \bibinfo{journal}{\emph{Neural computation}} \bibinfo{volume}{3}, \bibinfo{number}{1} (\bibinfo{year}{1991}), \bibinfo{pages}{79--87}.
\newblock


\bibitem[Jannach et~al\mbox{.}(2021)]%
        {jannach2021survey}
\bibfield{author}{\bibinfo{person}{Dietmar Jannach}, \bibinfo{person}{Ahtsham Manzoor}, \bibinfo{person}{Wanling Cai}, {and} \bibinfo{person}{Li Chen}.} \bibinfo{year}{2021}\natexlab{}.
\newblock \showarticletitle{A survey on conversational recommender systems}.
\newblock \bibinfo{journal}{\emph{ACM Computing Surveys (CSUR)}} \bibinfo{volume}{54}, \bibinfo{number}{5} (\bibinfo{year}{2021}), \bibinfo{pages}{1--36}.
\newblock


\bibitem[Lin et~al\mbox{.}(2025)]%
        {lin2025can}
\bibfield{author}{\bibinfo{person}{Jianghao Lin}, \bibinfo{person}{Xinyi Dai}, \bibinfo{person}{Yunjia Xi}, \bibinfo{person}{Weiwen Liu}, \bibinfo{person}{Bo Chen}, \bibinfo{person}{Hao Zhang}, \bibinfo{person}{Yong Liu}, \bibinfo{person}{Chuhan Wu}, \bibinfo{person}{Xiangyang Li}, \bibinfo{person}{Chenxu Zhu}, {et~al\mbox{.}}} \bibinfo{year}{2025}\natexlab{}.
\newblock \showarticletitle{How can recommender systems benefit from large language models: A survey}.
\newblock \bibinfo{journal}{\emph{ACM Transactions on Information Systems}} \bibinfo{volume}{43}, \bibinfo{number}{2} (\bibinfo{year}{2025}), \bibinfo{pages}{1--47}.
\newblock


\bibitem[Lonsdale and North(2011)]%
        {lonsdale2011we}
\bibfield{author}{\bibinfo{person}{Adam~J Lonsdale} {and} \bibinfo{person}{Adrian~C North}.} \bibinfo{year}{2011}\natexlab{}.
\newblock \showarticletitle{Why do we listen to music? A uses and gratifications analysis}.
\newblock \bibinfo{journal}{\emph{British Journal of Psychology}} \bibinfo{volume}{102}, \bibinfo{number}{1} (\bibinfo{year}{2011}), \bibinfo{pages}{108--134}.
\newblock


\bibitem[Loshchilov and Hutter(2016)]%
        {loshchilov2016sgdr}
\bibfield{author}{\bibinfo{person}{Ilya Loshchilov} {and} \bibinfo{person}{Frank Hutter}.} \bibinfo{year}{2016}\natexlab{}.
\newblock \showarticletitle{Sgdr: Stochastic gradient descent with warm restarts}.
\newblock \bibinfo{journal}{\emph{arXiv preprint arXiv:1608.03983}} (\bibinfo{year}{2016}).
\newblock


\bibitem[Loshchilov and Hutter(2019)]%
        {loshchilov2017decoupled}
\bibfield{author}{\bibinfo{person}{Ilya Loshchilov} {and} \bibinfo{person}{Frank Hutter}.} \bibinfo{year}{2019}\natexlab{}.
\newblock \showarticletitle{Decoupled weight decay regularization}. In \bibinfo{booktitle}{\emph{Proc. ICLR}}.
\newblock


\bibitem[Manco et~al\mbox{.}(2022)]%
        {manco2022contrastive}
\bibfield{author}{\bibinfo{person}{Ilaria Manco}, \bibinfo{person}{Emmanouil Benetos}, \bibinfo{person}{Elio Quinton}, {and} \bibinfo{person}{George Fazekas}.} \bibinfo{year}{2022}\natexlab{}.
\newblock \showarticletitle{Contrastive Audio-Language Learning for Music}. In \bibinfo{booktitle}{\emph{Ismir 2022 Hybrid Conference}}.
\newblock


\bibitem[Marey et~al\mbox{.}(2024)]%
        {marey2024modeling}
\bibfield{author}{\bibinfo{person}{Lilian Marey}, \bibinfo{person}{Bruno Sguerra}, {and} \bibinfo{person}{Manuel Moussallam}.} \bibinfo{year}{2024}\natexlab{}.
\newblock \showarticletitle{Modeling Activity-Driven Music Listening with PACE}. In \bibinfo{booktitle}{\emph{Proceedings of the 2024 Conference on Human Information Interaction and Retrieval}}. \bibinfo{pages}{346--351}.
\newblock


\bibitem[McFee et~al\mbox{.}(2012)]%
        {mcfee2012million}
\bibfield{author}{\bibinfo{person}{Brian McFee}, \bibinfo{person}{Thierry Bertin-Mahieux}, \bibinfo{person}{Daniel~PW Ellis}, {and} \bibinfo{person}{Gert~RG Lanckriet}.} \bibinfo{year}{2012}\natexlab{}.
\newblock \showarticletitle{The million song dataset challenge}. In \bibinfo{booktitle}{\emph{Proceedings of the 21st International Conference on World Wide Web}}. \bibinfo{pages}{909--916}.
\newblock


\bibitem[Meseguer-Brocal et~al\mbox{.}(2024)]%
        {meseguer2024experimental}
\bibfield{author}{\bibinfo{person}{Gabriel Meseguer-Brocal}, \bibinfo{person}{Dorian Desblancs}, {and} \bibinfo{person}{Romain Hennequin}.} \bibinfo{year}{2024}\natexlab{}.
\newblock \showarticletitle{An experimental comparison of multi-view self-supervised methods for music tagging}. In \bibinfo{booktitle}{\emph{ICASSP 2024-2024 IEEE International Conference on Acoustics, Speech and Signal Processing (ICASSP)}}. IEEE, \bibinfo{pages}{1141--1145}.
\newblock


\bibitem[Oramas et~al\mbox{.}(2024)]%
        {oramas2024talking}
\bibfield{author}{\bibinfo{person}{Sergio Oramas}, \bibinfo{person}{Andres Ferraro}, \bibinfo{person}{Alvaro Sarasua}, {and} \bibinfo{person}{Fabien Gouyon}.} \bibinfo{year}{2024}\natexlab{}.
\newblock \showarticletitle{Talking to Your Recs: Multimodal Embeddings For Recommendation and Retrieval}. In \bibinfo{booktitle}{\emph{Proceedings of the 2nd Music Recommender Systems Workshop 2024 co-located with the 18th ACM Conference on Recommender Systems (RecSys 2024)}}.
\newblock


\bibitem[Palumbo et~al\mbox{.}(2024)]%
        {palumbo2024text2tracks}
\bibfield{author}{\bibinfo{person}{Enrico Palumbo}, \bibinfo{person}{Gustavo Penha}, \bibinfo{person}{Andreas Damianou}, \bibinfo{person}{Jos{\'e} Luis~Redondo Garc{\'\i}a}, \bibinfo{person}{Timothy~Christopher Heath}, \bibinfo{person}{Alice Wang}, \bibinfo{person}{Hugues Bouchard}, {and} \bibinfo{person}{Mounia Lalmas}.} \bibinfo{year}{2024}\natexlab{}.
\newblock \showarticletitle{Text2Tracks: Generative Track Retrieval for Prompt-based Music Recommendation}. In \bibinfo{booktitle}{\emph{The 1st Workshop on Risks, Opportunities, and Evaluation of Generative Models in Recommender Systems (ROEGEN@RECSYS'24)}}.
\newblock


\bibitem[Rendle et~al\mbox{.}(2012)]%
        {rendle2012bpr}
\bibfield{author}{\bibinfo{person}{Steffen Rendle}, \bibinfo{person}{Christoph Freudenthaler}, \bibinfo{person}{Zeno Gantner}, {and} \bibinfo{person}{Lars Schmidt-Thieme}.} \bibinfo{year}{2012}\natexlab{}.
\newblock \showarticletitle{BPR: Bayesian personalized ranking from implicit feedback}.
\newblock \bibinfo{journal}{\emph{arXiv preprint arXiv:1205.2618}} (\bibinfo{year}{2012}).
\newblock


\bibitem[Shazeer et~al\mbox{.}(2017)]%
        {shazeer2017outrageously}
\bibfield{author}{\bibinfo{person}{Noam Shazeer}, \bibinfo{person}{Azalia Mirhoseini}, \bibinfo{person}{Krzysztof Maziarz}, \bibinfo{person}{Andy Davis}, \bibinfo{person}{Quoc Le}, \bibinfo{person}{Geoffrey Hinton}, {and} \bibinfo{person}{Jeff Dean}.} \bibinfo{year}{2017}\natexlab{}.
\newblock \showarticletitle{Outrageously large neural networks: The sparsely-gated mixture-of-experts layer}.
\newblock \bibinfo{journal}{\emph{arXiv preprint arXiv:1701.06538}} (\bibinfo{year}{2017}).
\newblock


\bibitem[Tekle et~al\mbox{.}(2024)]%
        {tekle2024music}
\bibfield{author}{\bibinfo{person}{Noah Tekle}, \bibinfo{person}{Alline Ayala}, \bibinfo{person}{Jonathan Haile}, \bibinfo{person}{Abdulla Alshabanah}, \bibinfo{person}{Corey Baker}, {and} \bibinfo{person}{Murali Annavaram}.} \bibinfo{year}{2024}\natexlab{}.
\newblock \showarticletitle{Music Recommendation through LLM Song Summary}. In \bibinfo{booktitle}{\emph{The 1st Workshop on Risks, Opportunities, and Evaluation of Generative Models in Recommender Systems (ROEGEN@RECSYS'24)}}.
\newblock


\bibitem[Touvron et~al\mbox{.}(2023)]%
        {touvron2023llama}
\bibfield{author}{\bibinfo{person}{Hugo Touvron}, \bibinfo{person}{Thibaut Lavril}, \bibinfo{person}{Gautier Izacard}, \bibinfo{person}{Xavier Martinet}, \bibinfo{person}{Marie-Anne Lachaux}, \bibinfo{person}{Timoth{\'e}e Lacroix}, \bibinfo{person}{Baptiste Rozi{\`e}re}, \bibinfo{person}{Naman Goyal}, \bibinfo{person}{Eric Hambro}, \bibinfo{person}{Faisal Azhar}, {et~al\mbox{.}}} \bibinfo{year}{2023}\natexlab{}.
\newblock \showarticletitle{Llama: Open and efficient foundation language models}.
\newblock \bibinfo{journal}{\emph{arXiv preprint arXiv:2302.13971}} (\bibinfo{year}{2023}).
\newblock


\bibitem[Tran et~al\mbox{.}(2021)]%
        {tran2021hierarchical}
\bibfield{author}{\bibinfo{person}{Viet-Anh Tran}, \bibinfo{person}{Guillaume Salha-Galvan}, \bibinfo{person}{Romain Hennequin}, {and} \bibinfo{person}{Manuel Moussallam}.} \bibinfo{year}{2021}\natexlab{}.
\newblock \showarticletitle{Hierarchical latent relation modeling for collaborative metric learning}. In \bibinfo{booktitle}{\emph{Proceedings of the 15th ACM Conference on Recommender Systems}}. \bibinfo{pages}{302--309}.
\newblock


\bibitem[Tran et~al\mbox{.}(2024)]%
        {tran2024transformers}
\bibfield{author}{\bibinfo{person}{Viet-Anh Tran}, \bibinfo{person}{Guillaume Salha-Galvan}, \bibinfo{person}{Bruno Sguerra}, {and} \bibinfo{person}{Romain Hennequin}.} \bibinfo{year}{2024}\natexlab{}.
\newblock \showarticletitle{Transformers Meet ACT-R: Repeat-Aware and Sequential Listening Session Recommendation}. In \bibinfo{booktitle}{\emph{Proceedings of the 18th ACM Conference on Recommender Systems}}. \bibinfo{pages}{486--496}.
\newblock


\bibitem[Turrin et~al\mbox{.}(2015)]%
        {DBLP:conf/recsys/TurrinQCPC15}
\bibfield{author}{\bibinfo{person}{Roberto Turrin}, \bibinfo{person}{Massimo Quadrana}, \bibinfo{person}{Andrea Condorelli}, \bibinfo{person}{Roberto Pagano}, {and} \bibinfo{person}{Paolo Cremonesi}.} \bibinfo{year}{2015}\natexlab{}.
\newblock \showarticletitle{30Music Listening and Playlists Dataset}. In \bibinfo{booktitle}{\emph{Poster Proceedings of the 9th {ACM} Conference on Recommender Systems, RecSys 2015, Vienna, Austria, September 16, 2015}} \emph{(\bibinfo{series}{{CEUR} Workshop Proceedings}, Vol.~\bibinfo{volume}{1441})}. \bibinfo{publisher}{CEUR-WS.org}.
\newblock
\urldef\tempurl%
\url{https://ceur-ws.org/Vol-1441/recsys2015\_poster13.pdf}
\showURL{%
\tempurl}


\bibitem[Vaswani et~al\mbox{.}(2017)]%
        {vaswani2017attention}
\bibfield{author}{\bibinfo{person}{Ashish Vaswani}, \bibinfo{person}{Noam Shazeer}, \bibinfo{person}{Niki Parmar}, \bibinfo{person}{Jakob Uszkoreit}, \bibinfo{person}{Llion Jones}, \bibinfo{person}{Aidan~N Gomez}, \bibinfo{person}{{\L}ukasz Kaiser}, {and} \bibinfo{person}{Illia Polosukhin}.} \bibinfo{year}{2017}\natexlab{}.
\newblock \showarticletitle{Attention is all you need}.
\newblock \bibinfo{journal}{\emph{Advances in neural information processing systems}}  \bibinfo{volume}{30} (\bibinfo{year}{2017}).
\newblock


\bibitem[Wang et~al\mbox{.}(2024)]%
        {wang2024multilingual}
\bibfield{author}{\bibinfo{person}{Liang Wang}, \bibinfo{person}{Nan Yang}, \bibinfo{person}{Xiaolong Huang}, \bibinfo{person}{Linjun Yang}, \bibinfo{person}{Rangan Majumder}, {and} \bibinfo{person}{Furu Wei}.} \bibinfo{year}{2024}\natexlab{}.
\newblock \showarticletitle{Multilingual e5 text embeddings: A technical report}.
\newblock \bibinfo{journal}{\emph{arXiv preprint arXiv:2402.05672}} (\bibinfo{year}{2024}).
\newblock


\bibitem[Warner et~al\mbox{.}(2024)]%
        {warner2024smarter}
\bibfield{author}{\bibinfo{person}{Benjamin Warner}, \bibinfo{person}{Antoine Chaffin}, \bibinfo{person}{Benjamin Clavi{\'e}}, \bibinfo{person}{Orion Weller}, \bibinfo{person}{Oskar Hallstr{\"o}m}, \bibinfo{person}{Said Taghadouini}, \bibinfo{person}{Alexis Gallagher}, \bibinfo{person}{Raja Biswas}, \bibinfo{person}{Faisal Ladhak}, \bibinfo{person}{Tom Aarsen}, {et~al\mbox{.}}} \bibinfo{year}{2024}\natexlab{}.
\newblock \showarticletitle{Smarter, better, faster, longer: A modern bidirectional encoder for fast, memory efficient, and long context finetuning and inference}.
\newblock \bibinfo{journal}{\emph{arXiv preprint arXiv:2412.13663}} (\bibinfo{year}{2024}).
\newblock


\bibitem[Xu et~al\mbox{.}(2024)]%
        {xu2024prompting}
\bibfield{author}{\bibinfo{person}{Lanling Xu}, \bibinfo{person}{Junjie Zhang}, \bibinfo{person}{Bingqian Li}, \bibinfo{person}{Jinpeng Wang}, \bibinfo{person}{Mingchen Cai}, \bibinfo{person}{Wayne~Xin Zhao}, {and} \bibinfo{person}{Ji-Rong Wen}.} \bibinfo{year}{2024}\natexlab{}.
\newblock \showarticletitle{Prompting large language models for recommender systems: A comprehensive framework and empirical analysis}.
\newblock \bibinfo{journal}{\emph{arXiv preprint arXiv:2401.04997}} (\bibinfo{year}{2024}).
\newblock


\bibitem[Zhao et~al\mbox{.}(2024)]%
        {zhao2024recommender}
\bibfield{author}{\bibinfo{person}{Zihuai Zhao}, \bibinfo{person}{Wenqi Fan}, \bibinfo{person}{Jiatong Li}, \bibinfo{person}{Yunqing Liu}, \bibinfo{person}{Xiaowei Mei}, \bibinfo{person}{Yiqi Wang}, \bibinfo{person}{Zhen Wen}, \bibinfo{person}{Fei Wang}, \bibinfo{person}{Xiangyu Zhao}, \bibinfo{person}{Jiliang Tang}, {et~al\mbox{.}}} \bibinfo{year}{2024}\natexlab{}.
\newblock \showarticletitle{Recommender systems in the era of large language models (llms)}.
\newblock \bibinfo{journal}{\emph{IEEE Transactions on Knowledge and Data Engineering}} (\bibinfo{year}{2024}).
\newblock


\end{thebibliography}

\appendix
\end{document}